\documentclass[twocolumn]{aastex701}
\usepackage{amsmath}
\usepackage{svg}
\usepackage{nicefrac}
\usepackage[none]{hyphenat}

\def\Msun{M$_{\odot}$}

\begin{document}

\title{New Insights into the T Tauri Binary Separation Distribution} 

\author{Caleb Eastlund}
\affiliation{Department of Physics and Astronomy, University~of~Wyoming, 1000~E.~University~Ave., Dept.~3905, Laramie, WY 82071, USA}
\email{ceastlun@uwyo.edu}

\author{Maxwell~Moe}
\affiliation{Department of Physics and Astronomy, University~of~Wyoming, 1000~E.~University~Ave., Dept.~3905, Laramie, WY 82071, USA}
\email{mmoe2@uwyo.edu}

\author{Kaitlin~M.~Kratter}
\affiliation{Steward Observatory, University of Arizona, 933~N.~Cherry~Ave.,~Tucson,~AZ 85721,~USA}
\email{kkratter@arizona.edu}

\author{Marina Kounkel}
\affiliation{Department of Physics and Astronomy, University~of~North~Florida, 1~UNF~Drive, Building 50/2600, Jacksonville, FL 32224, USA}
\email{marina.kounkel@unf.edu}




\begin{abstract}
For three decades, adaptive optic surveys have revealed an excess of T~Tauri binaries across $a$~=~10\,-\,100~au in nearby star-forming regions compared to the field population of main-sequence (MS) stars. Such an excess requires that most stars are born in dense clusters and subjected to significant dynamical processing that disrupts such binaries across intermediate separations. However, we demonstrate that the apparent excess is due to an observational selection bias. Close binaries within $a$~$<$~100~au clear out their dusty circumstellar disks on faster timescales compared to wide binaries and single stars. A magnitude-limited sample is therefore biased toward close binaries that have preferentially cleared out their obscuring disks. We re-examine the separation distribution of pre-MS binaries in low-density Taurus, moderately dense Upper Scorpius, and the extremely dense Orion Nebula Cluster (ONC). By limiting the samples to primary spectral type / mass instead of magnitude, the artificial excess across $a$ = 10\,-\,100 au disappears in all three environments. Across wider separations $a$ = 100\,-\,4,000 au, Taurus exhibits an excess of companions (mostly tertiaries), the ONC displays a deficit, and Upper Scorpius matches the field MS population. The field derives from an amalgam of all three environments, where Upper Scorpius corresponds to the average birth environment of solar-type stars. The total binary fraction within $a$~$<$~10,000~au in Taurus is only 52\%\,$\pm$\,7\%, substantially lower than the 100\% inferred from the biased observations and only slightly higher than the field MS value of 45\%. N-body interactions preferentially disrupt outer tertiaries with only marginal dynamical processing of the inner binaries, especially those within $a$ $<$ 100 au. 
\end{abstract}


\section{Introduction}
\label{sec:Intro}

Most solar-type stars are born in binary or multiple systems (see \citealt{Offner2023} for a recent review). During the first $\approx$\,0.5 Myr (embedded Class 0/I phase), dynamical friction with the surrounding gas re-shapes the binary separation distribution \citep{Bate2003,Bate2009,Lee2019,TokovininMoe2020}. On longer timescales (Class II/III T~Tauri phase), N-body interactions with other stars in their birth cluster further evolve the orbital architectures, typically disrupting the widest systems \citep{Heggie1975,Kroupa1995,Moe2018,CournoyerCloutier2024}. 

For three decades, adaptive optics (AO) and speckle imaging surveys have revealed a factor of $\approx$\,2 excess of Class II/III T~Tauri binaries across separations $a$~=~10\,-\,100~au in nearby star-forming regions (SFRs) compared to the field population of main-sequence (MS) stars \citep{Ghez1993,Leinert1993,Ghez1997,Duchene2007,Kraus2011,Tokovinin2020}. The cited surveys focused on SFRs with low to moderate stellar densities such as Taurus-Auriga (hereafter Taurus), Ophiuchus, Chamaeleon, and Scorpius-Centaurus, which includes the three sub-groups Upper Scorpius (hereafter Upper Sco), Upper Centaurus-Lupus, and Lower Centaurus-Crux. It was originally believed that only low-density SFRs would exhibit such a binary excess across 10\,-\,100~au and that high-density environments should instead exhibit a deficit due to dynamical disruptions. 

However, recent observations have called the T~Tauri binary excess across $a$~=~10\,-\,100~au into question \citep[][see details below]{Duchene2018,Kounkel2019}. Moreover, the magnitude-limited imaging surveys are prone to a dust-extinction selection bias not yet accounted for in the previous studies. In Section~\ref{sec:Motivation}, we outline these inconsistencies and motivate the need to account for dust extinction from circumstellar disks. In Section~\ref{sec:Previous}, we summarize the methods and results of previous AO/speckle imaging surveys of T Tauri binaries. In Section~\ref{sec:Impact}, we illustrate how dust extinction biases the previously measured T Tauri binary fractions and separation distributions. We present our bias-corrected results in Section~\ref{sec:Results} and compare the different SFRs in  Section~\ref{sec:Comparison}. We discuss the implications for N-body interactions in Section ~\ref{sec:Implications} and summarize our key results in Section~\ref{sec:Summary}.

\section{Motivation}
\label{sec:Motivation}

\subsection{Two Inconsistencies}

\begin{figure}[t!]
    \centering
    \includegraphics[trim = 3mm 5mm 3mm 5mm, width=3.4in]{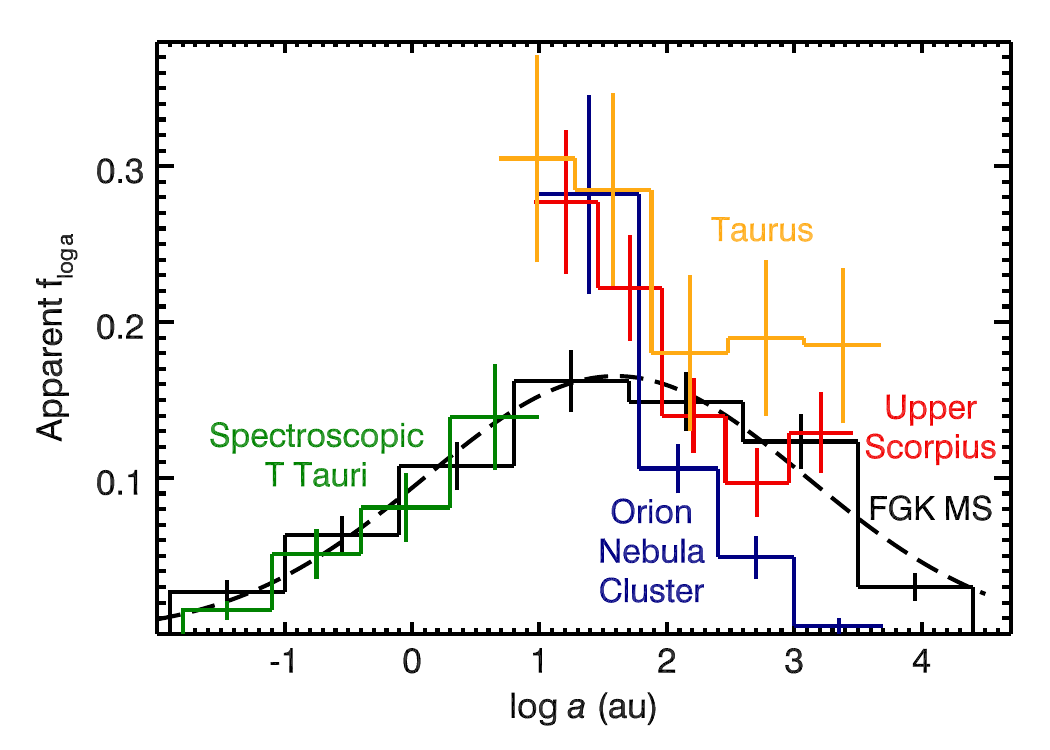}
    \caption{The spectroscopic T Tauri binary fraction within $a$~$<$~10~au \citep[green;][]{Kounkel2019} matches the field population of FGK MS binaries \citep[black;][]{Raghavan2010}. However, sparse Taurus \citep[orange;][]{Kraus2011}, moderately dense Upper Sco \citep[red;][]{Tokovinin2020}, and even the extremely dense ONC \citep[blue;][]{Duchene2018} all exhibit an apparent excess of companions across $a$ = 10\,-\,60~au, which we attribute to a dust-extinction selection bias in the present study.}
    \label{fig:All_Before}
\end{figure}

In Fig.~\ref{fig:All_Before}, we display the companion fraction  per decade of orbital separation $f_{\rm loga}$ adapted from \citet{Offner2023} for Taurus \citep{Kraus2011}, Upper~Sco \citep{Tokovinin2020}, and the field FGK-type MS population \citep{Raghavan2010}. Both Taurus and Upper Sco exhibit an excess of binaries across $a$ = 10\,-100~au compared to the field MS population. This apparent excess is inconsistent with two recent observations.

First, the completeness-corrected spectroscopic binary fraction and period distribution of T Tauri stars match the field MS population with no evidence for evolution with cluster age or density \citep{Mathieu1994,Melo2003,Prato2007,Kounkel2019}.  In particular, \citet{Kounkel2019} utilized multi-epoch near-infrared SDSS-APOGEE spectra to recover the intrinsic close binary fraction and separation distribution within $a$~$<$~10~au. Their survey included sparse Taurus, dense ONC, and even slightly older clusters like the Pleiades. \citet{Kounkel2019} measured a T~Tauri close binary fraction that increases with mass and matches the field MS population (see their Fig.~16). However, they did not find any statistically significant variations with respect to environment (see their Fig.~17). As shown in Fig.~\ref{fig:All_Before}, there is a factor of two discontinuity at $a$~=~10~au whereby spectroscopic surveys find a T Tauri close binary fraction that is consistent with the field population while imaging surveys reveal a factor of two excess across all SFRs.

Second, AO imaging at VLT revealed a factor of $\approx$\,2 excess of solar-type T~Tauri binaries across $a$~=~10\,-\,60~au in the dense Orion Nebula Cluster \citep[ONC;][]{Duchene2018}. Previous studies showed that T Tauri stars in the ONC have a deficit of wider companions across $a$~=~100\,-\,10,000~au due to dynamical disruptions \citep{Scally1999,Reipurth2007}, but the excess across 10\,-\,60 au was surprising (see Fig.~\ref{fig:All_Before}). Such binaries at intermediate separations will remain gravitationally bound as the ONC continues to dissolve. \citet{Duchene2018} concluded, ``If most stars in the field arise from regions similar to, or less dense than, the ONC, they would host a higher frequency of close visual binaries. This may indicate that nearby SFRs are not representative of the conditions that reigned when the majority of field stars formed, several Gyr ago.'' Utilizing archival HST observations of the ONC, \citet{DeFurio2019} subsequently measured a binary fraction across 10\,-\,200~au of low-mass stars that matches their field M-dwarf counterparts. Nonetheless, we have yet to identify a SFR that exhibits a deficit of solar-type binaries across $a$~=~10\,-\,60~au. If the field population derives from a distribution of cluster masses and densities, then for every solar-type star born in Taurus, Upper Sco, and possibly Orion with an excess of companions across 10\,-\,60~au, then there must be a corresponding solar-type star born in an extremely dense environment, even denser than the ONC, with a counteracting deficit.  However, observations and dynamical modeling of open clusters in our Milky Way and nearby galaxies suggest that the ONC already corresponds to the dense, high-mass end of the initial cluster mass function \citep{Lamers2003,Gieles2006,Piskunov2008,Fujii2016,Krumholz2019}. It is thus difficult to conclude that half of field solar-type stars were born in SFRs even denser than the ONC.

\subsection{Selection Bias from Dusty Circumstellar Disks}

The goal of the present study is to critically examine the apparent excess of T~Tauri binaries measured by previous imaging surveys of Taurus, Upper Sco, and the ONC. We focus on a particular selection bias not previously considered whereby dust extinction from the surrounding circumstellar disks and envelopes strongly bias a magnitude-limited sample. Binary stars within $a$~$<$~100 au truncate and clear out their protostellar disks on faster timescales compared to their wide binary or single star counterparts \citep{Jensen1994,Jensen1996, Kraus2012,Harris2012,Cheetham2015}. For 2-Myr-old Taurus members, \citet{Kraus2012} demonstrated that 80\% of single stars and 90\% of wide binaries beyond $a$~$>$~40~au have disks while only 37\% of close visual binaries across $a$ = 2\,-\,40~au still retain their disks. The imaging survey samples, which were mostly selected according to their optical magnitudes, could potentially be biased toward close binaries that have preferentially cleared out their disks and experience less dust extinction. In contrast, the spectroscopic binary samples were largely based on spectral type or near-IR magnitudes where dust extinction is mostly negligible. The T Tauri spectroscopic binary fraction measured by \citet{Kounkel2019} is thus not prone to this dust-extinction selection bias.

\begin{figure}[t!]
    \centering
    \includegraphics[trim = 5mm 5mm 3mm 5mm, width=3.4in]{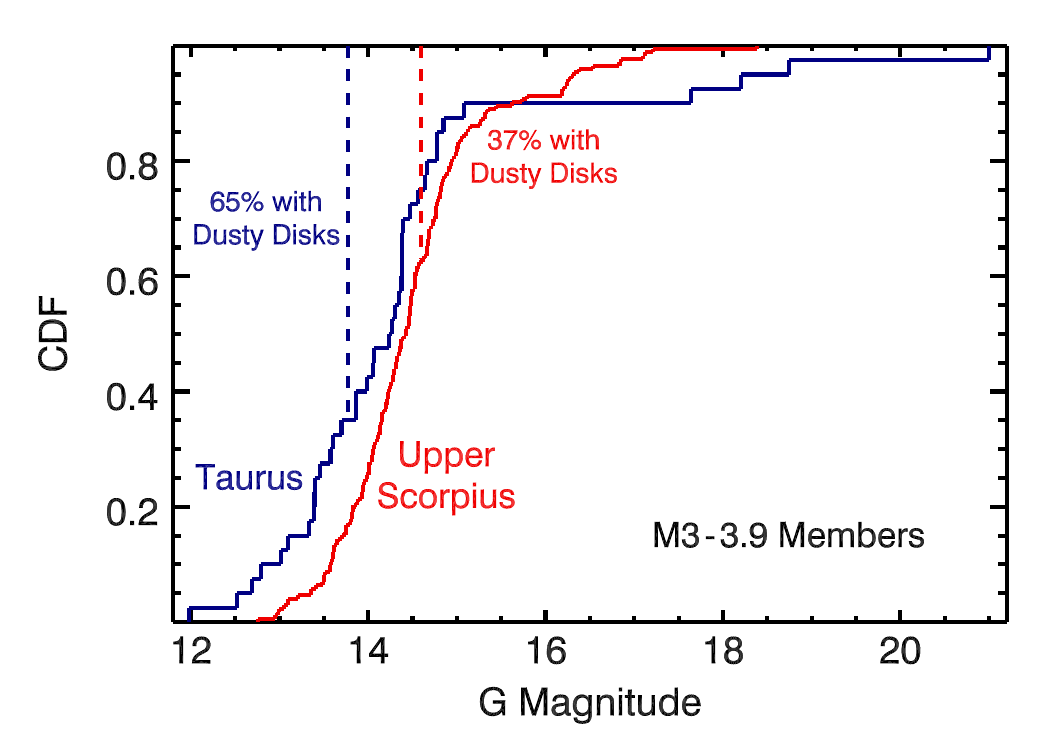}
    \caption{Cumulative distribution functions of G magnitudes for M3-M3.9 members in Taurus (blue) and Upper Sco (red). Stars fainter than  $\Delta$G $>$ 1.83 mag of the brightest members (right of dashed lines) are embedded in circumstellar disks and suffer from non-negligible dust extinction. Magnitude-limited samples are biased against stars that have retained their dusty disks.}
    \label{fig:dustydisk}
\end{figure}

To illustrate the significance of dust extinction from circumstellar disks/envelopes, we display in Fig.~\ref{fig:dustydisk} the cumulative distribution functions (CDFs) of G magnitudes for all M3-3.9 members in both 2-Myr-old Taurus \citep{Luhman2023} and 11-Myr-old Upper Sco \citep{Esplin2018}. The expected width of zero-age MS stars with M3-M3.9 spectral types is only  $\Delta$G = 1.08~mag \citep{Pecaut2013}. Twin binaries can potentially be  $\Delta$G = 0.75~mag brighter than their single-star counterparts. In Fig.~\ref{fig:dustydisk}, we display vertical dashed lines at $\Delta$G = 1.08 + 0.75 = 1.83~mag fainter than the brightest M3-M3.9 member in each SFR. Stars to the right of these dashed lines must be embedded in circumstellar disks and suffer a non-negligible amount of dust extinction. In Taurus, we estimate that 65\% of M3-3.9 members are embedded in dusty disks, some with up to A$_{\rm G}$ = 7 mag of dust extinction. Older M3-M3.9 stars in Upper Sco experience less dust extinction from circumstellar disks, only up to A$_{\rm G}$ = 4 mag. Nonetheless, a non-negligible 37\% of M3-M3.9 Upper Sco members still retain their obscuring disks. Magnitude-limited samples are biased against these members that are still embedded in dusty disks.

\section{Summary of Previous Observations}
\label{sec:Previous}

\subsection{Taurus}

\begin{figure*}[t!]
    \centering
    \includegraphics[scale=0.65]{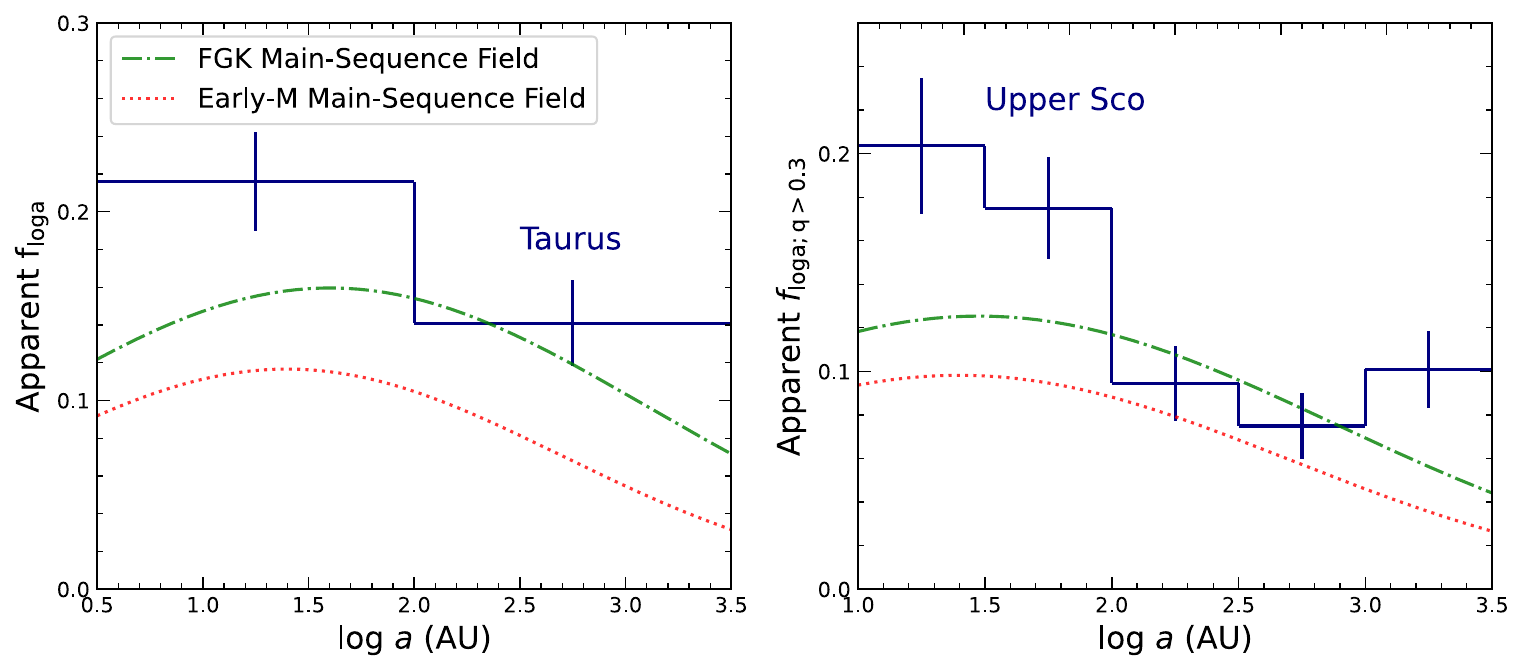}
    \caption{The apparent binary separation distribution of confirmed Taurus members from \citet[][left panel]{Kraus2011} and confirmed Upper Sco members from \citet[][right panel]{Tokovinin2020}. We display the corresponding FGK (green) and early-M (red) MS field distributions.  Before accounting for the dust-extinction selection bias, both Taurus and Upper Sco exhibit an apparent excess of close binaries within $a$ $<$ 100 au.}
    \label{fig:SepDist_NoCut}
\end{figure*}

We first summarize the properties of the different star-forming regions and the observations for visual companions to their members. At a distance of 145~pc, the Taurus-Auriga complex is a low-density SFR containing young stellar objects that are only 1\,-\,3~Myr old \citep{Kenyon2008,Torres2009,Dzib2015,Galli2018}. \citet{Luhman2023} identified 532 members based largely on the precise parallaxes and proper motions from {\it Gaia} DR3 \citep{Gaia2023}. \citet{Luhman2023} also listed their adopted spectral types for 497 members according to 84 different spectroscopic surveys and references. Most of the remaining 35 members without spectral types are faint or highly embedded infrared sources with only IRAS designations. 

\citet{Kraus2011} targeted 82 probable and 10 candidate Taurus members with near-IR AO at both Keck and Palomar. They discovered 16 new companions within $<$\,0.3'' of their primary stars. Their full statistical sample includes 60 additional likely Taurus members that were previously observed, including 44 with known companions. Of the 152 stars in the \citet{Kraus2011} sample, we remove the 10 non-members according to the \citet{Luhman2023} classification, leaving 142 confirmed Taurus members, all with known spectral types.  We adopt the \citet{Kraus2011} conversion between primary mass and spectral type, and we use their binary mass ratios $q$ = $M_2$/$M_1$ measured from the observed near-IR brightness contrasts. The \citet{Kraus2011} AO survey was sensitive to all stellar-mass companions with $M_2$~$\ge$~0.08\,\Msun\ beyond $a$~$>$~3\,AU (see their Fig. 3). In fact, \citet{Kraus2011} identified six brown dwarf companions below $M_2$~$<$~0.08\,\Msun, which we exclude in our statistical analysis and comparisons with the field binary star populations.

In the left panel of Fig.~\ref{fig:SepDist_NoCut}, we display the binary separation distribution for the full sample of confirmed Taurus members with AO observations from \citet{Kraus2011}. We overlay the canonical log-normal binary separation distributions \citep[Table 2 in][]{Offner2023} for field solar-type dwarfs ($\mu$~=~40~au, $\sigma_{\rm log\,a}$~=~1.5, companion frequency CF~=~0.60) and early-M dwarfs ($\mu$~=~25~au, $\sigma_{\rm log\,a}$~=~1.3, CF~=~0.38). In Table~\ref{tab:Taurus}, we list the properties of these samples and their binary fractions across $a$ = 3\,-\,100 au. The stellar companion fraction in Taurus is 32\%\,$\pm$\,4\% across $a$~=~3\,-\,100~au, a factor of 1.6 excess compared to field MS stars of similar mass that is statistically significant at the 3.3$\sigma$ level.

\begin{deluxetable}{lcc}[t!]
    \tablecaption{Binary Properties in Taurus versus the Field}\label{tab:Taurus}
    \startdata
      & & \\
      & & Binary Fraction  \\
     Sample &  $\langle M_1 \rangle$ & (a = 3 - 100 au) \\
     \hline
     \rule{0pt}{3ex}Field Early M-dwarfs & 0.4\,\Msun\ & 17\% \\
     \rule{0pt}{3ex}Field FGK-dwarfs & 0.9\,\Msun\ & 23\% \\
     \rule{0pt}{3ex}Taurus: Full Sample & 0.7\,\Msun\ & 32\%\,$\pm$\,4\% \\
     \rule{0pt}{3ex}Taurus: G0\,-\,M1.9 only & 0.8\,\Msun\ & 27\%\,$\pm$\,5\% \\
    \enddata
    
\end{deluxetable}

\subsection{Upper Scorpius}

Compared to Taurus, Upper Scorpius is at a similar distance of 145~pc, moderately denser, and substantially older at an age of 11 Myr \citep{Luhman2018}. \citet{Esplin2018} and \citet{Luhman2018} identified 1,608 Upper Sco members through a combination of mid-IR WISE photometry and {\it Gaia} DR1 astrometry. \citet{Esplin2018} also listed spectral types for each of these members. 

\citet{Tokovinin2020} performed a speckle-interferometric survey of the 614 Upper~Sco members brighter than I $<$ 13, including 10 wide companions. Their sample consists of the confirmed Upper Sco members from \citet{Esplin2018} and \citet{Luhman2018}, but they also used {\it Gaia} DR2 to further vet their sample. \citet{Tokovinin2020} identified 187 pairs via speckle interferometry, including 55 new discoveries. They complimented their own observations with {\it Gaia} DR2 common-proper-motion pairs, resulting in a total of 250 binaries among 604 primaries.  \citet{Tokovinin2020} accounted for the \citet{Branch1976} bias by removing the 10 binaries with photometric primary masses below $M_1$ $<$ 0.38\,\Msun\ (see their Fig.~5). The \citet{Tokovinin2020} speckle-interferometric survey was complete to nearly all $q$~$>$~0.3 companions beyond $a$~$>$~100~au (see their Fig.~15). At shorter separations, we adopt their detection efficiency of $q$~$>$~0.3 companions, which for their full sample increases from 64\% across log\,$a$\,(au) = 1.0\,-\,1.5 to 95\% across log\,$a$\,(au) = 1.5\,-\,2.0 (see their Table~7). 

We display in the right panel of Fig.~\ref{fig:SepDist_NoCut} the completeness-corrected separation distribution $f_{\rm loga;q>0.3}$ of companions with $q$~$>$~0.3 for all Upper~Sco primaries observed by \citet{Tokovinin2020}. For field early-M dwarf binaries, the log-normal separation distribution of companions with $q$~$>$~0.3 follows the same form as those with $M_2$~$>$~0.08\,\Msun\ ($\mu$ = 25 au, $\sigma_{\rm log\,a}$ = 1.3), but normalized to a slightly smaller companion frequency (CF = 0.32). Meanwhile, solar-type binaries are weighted toward larger mass ratios at closer separations \citep{Moe2017}, and the solar-type distribution of companions with $q$~$>$~0.3 shifts accordingly ($\mu$ = 30 au, $\sigma_{\rm log\,a}$ = 1.4, CF = 0.44). We overlay both early-M and solar-type field distributions of $f_{\rm loga;q>0.3}$ in the right panel of Fig.~\ref{fig:SepDist_NoCut}. In Table~\ref{tab:UpperSco}, we list the binary fractions across $a$ = 10\,-\,100 au and above $q$ $>$ 0.3 for each of these samples. The completeness-corrected binary fraction across $a$ = 10\,-\,100~au and above $q$ $>$ 0.3 for the full Upper Sco sample is 19\%\,$\pm$\,2\%, which is 1.6 times the corresponding field MS value at the 3.6$\sigma$ significance level. 

\begin{deluxetable}{lcc}[t!]
    \tablecaption{Binary Properties in Upper Sco versus the Field}\label{tab:UpperSco}
    \startdata
      & & \\
      & & Binary Fraction  \\
      & & ($a$ = 10\,-\,100 au; \\
     Sample &  $\langle M_1 \rangle$ & $q$ $>$ 0.3) \\
     \hline
     \rule{0pt}{3ex}Field Early M-dwarfs & 0.4\,\Msun\ & 10\% \\
     \rule{0pt}{3ex}Field FGK-dwarfs & 0.9\,\Msun\ & 13\% \\
     \rule{0pt}{3ex}Upper Sco: Full Sample & 0.8\,\Msun\ & 19\%\,$\pm$\,2\% \\
     \rule{0pt}{3ex}Upper Sco: A6\,-\,M2.9 only & 0.9\,\Msun\ & 15\%\,$\pm$\,2\% \\
    \enddata
\end{deluxetable}

\subsection{Orion Nebula Cluster}

At only $\approx$\,1 Myr old, the ONC is the youngest and densest star-forming region considered in this study \citep{Hillenbrand1997,DaRio2010,Allen2017,Kounkel2018}. \citet{Duchene2018} performed a near-IR AO survey with the NACO instrument at the VLT for a magnitude-limited sample of 42 ONC members across K = 7.5\,-\,9.5~mag. They were relatively complete toward stellar-mass companions across 26\,-\,155~mas ($a$ = 10\,-\,60~au given an ONC distance of 390 pc), where they resolved 12 companions. \citet{Duchene2018} accounted for the \citet{Branch1976} bias by removing the four tight visual binaries with primary stars that were fainter than K~$>$~9.5 after deducting the flux from the resolved companions. We summarize the binary properties of ONC members in Table~\ref{tab:Orion}. The measured companion fraction of 21$_{-5}^{+8}$\% across $a$ = 10\,-\,60 au is 1.7 times larger than the field solar-type MS value of 13\%, statistically significant at the 1.7$\sigma$ level. 

\begin{deluxetable}{lcc}[t!]
    \tablecaption{Binary Properties in the ONC versus the Field} \label{tab:Orion}
    \startdata
      & & \\
      & & Binary Fraction  \\
     Sample &  $\langle M_1 \rangle$ & ($a$ = 10\,-\,60~au) \\
     \hline
     \rule{0pt}{3ex}Field FGK-dwarfs & 0.9\,\Msun\ & 13\% \\
     \rule{0pt}{3ex}ONC: Full Sample & 0.9\,\Msun\ & 21$_{-5}^{+8}$\% \\
     \rule{0pt}{3ex}ONC: $M_1$ = 0.7\,-\,1.6\,\Msun\ only & 1.1\,\Msun\ & 19$_{-7}^{+12}$\% \\
    \enddata
\end{deluxetable}

\section{Impact of the Dust-Extinction Selection Bias}
\label{sec:Impact}

\begin{figure*}[t!]
    \centering
    \includegraphics[scale=0.65]{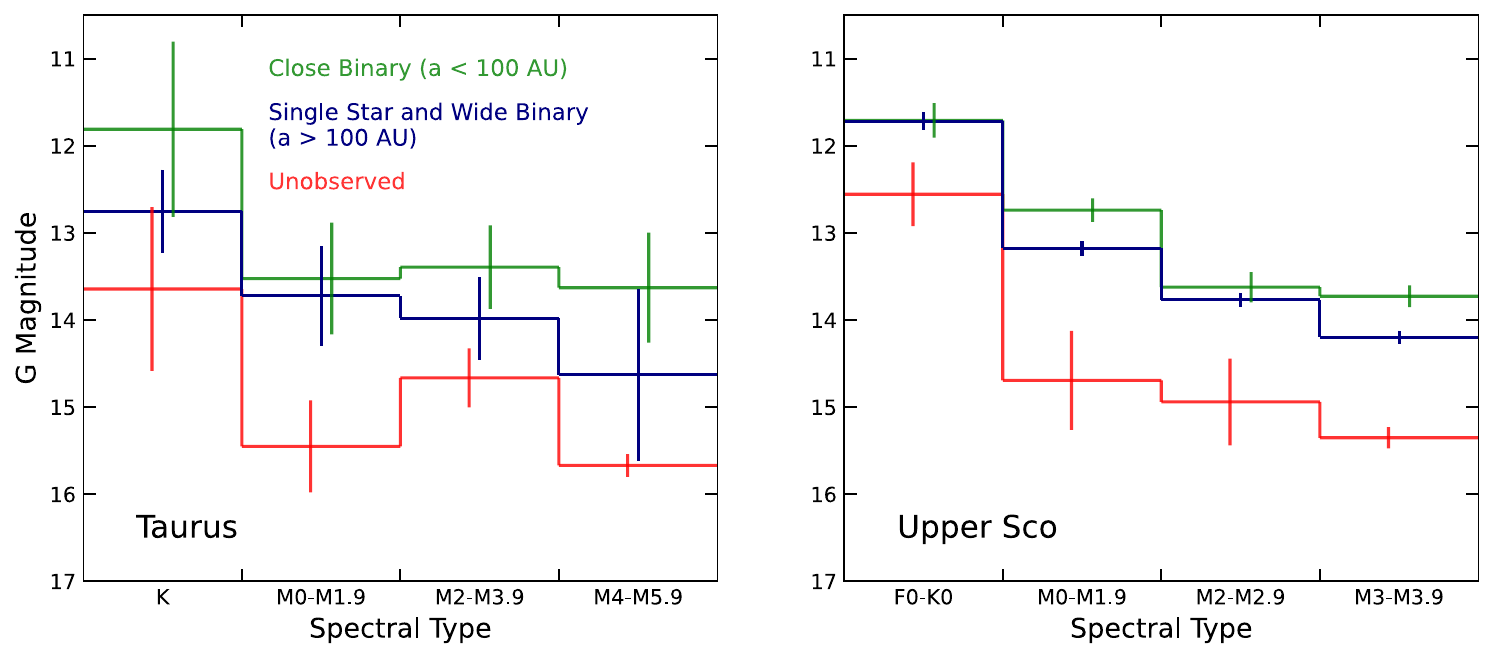}
    \caption{Average G magnitudes of Taurus members (left panel) and Upper Sco members (right panel) as a function of spectral type for three populations: close binaries with $a$~$\le$~100~au (green), single stars and wide binaries with $a$ $>$ 100~au (blue), and members that were not targeted by AO/speckle imaging (red). Close binaries are systematically brighter, which cannot be explained by the \citet{Branch1976} bias but instead is due to the preferential clearing of their dusty, obscuring disks. }
    \label{fig:MeanG}
\end{figure*}

We now investigate the impact of the dust-extinction selection bias on the observed samples. In Fig.~\ref{fig:MeanG}, we show the average G magnitudes as a function of primary spectral type for three populations: close binaries within $a$~$\le$~100~au, single stars and wide binaries beyond $a$~$>$~100~au, and members that were not targeted by AO/speckle imaging for visual binaries. The average brightness decreases toward later spectral types as expected. For a given primary spectral type, close binaries are brighter than single stars / wide binaries, which in turn are brighter than the unobserved members. A magnitude-limited survey is slightly biased toward twin binaries \citep{Branch1976}. Nonetheless, twin binaries with $q$ $>$ 0.8 are only 0.4\,-\,0.7~mag brighter than their single-star counterparts. Even after removing twin binaries with $q$ $>$ 0.8, close binaries are still brighter than the unobserved systems at a statistically significant level. We conclude that close binaries are brighter because they have cleared out their dusty disks. We surmise that the fainter unobserved systems are dominated by highly obscured single stars and wide binaries.

\begin{figure*}[t!]
    \centering
    \includegraphics[scale=0.65]{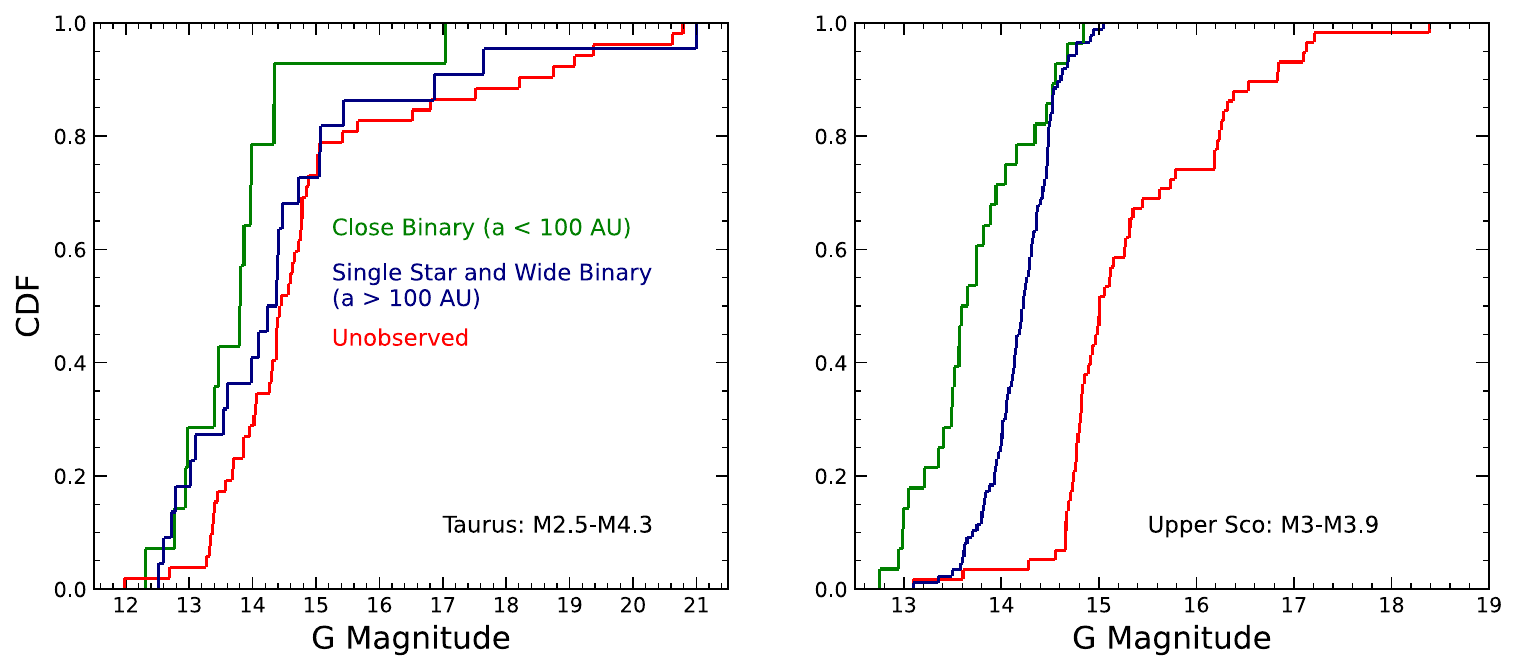}
    \caption{Cumulative distribution functions of G~magnitudes for the 88 M2.5\,-\,M4.3 Taurus members (left panel) and 173 M3\,-\,3.9 Upper~Sco members (right panel) separated into the same three populations as in Fig.~\ref{fig:MeanG}. For a narrow range of spectral types, close binaries are narrowly distributed in brightness because they have preferentially cleared out their disks. Meanwhile, a larger fraction of single stars, wide binaries, and unobserved members retain their dusty obscuring disks.}
    \label{fig:CDF}
\end{figure*}

\begin{figure*}[t!]
    \centering
    \includegraphics[scale=0.67]{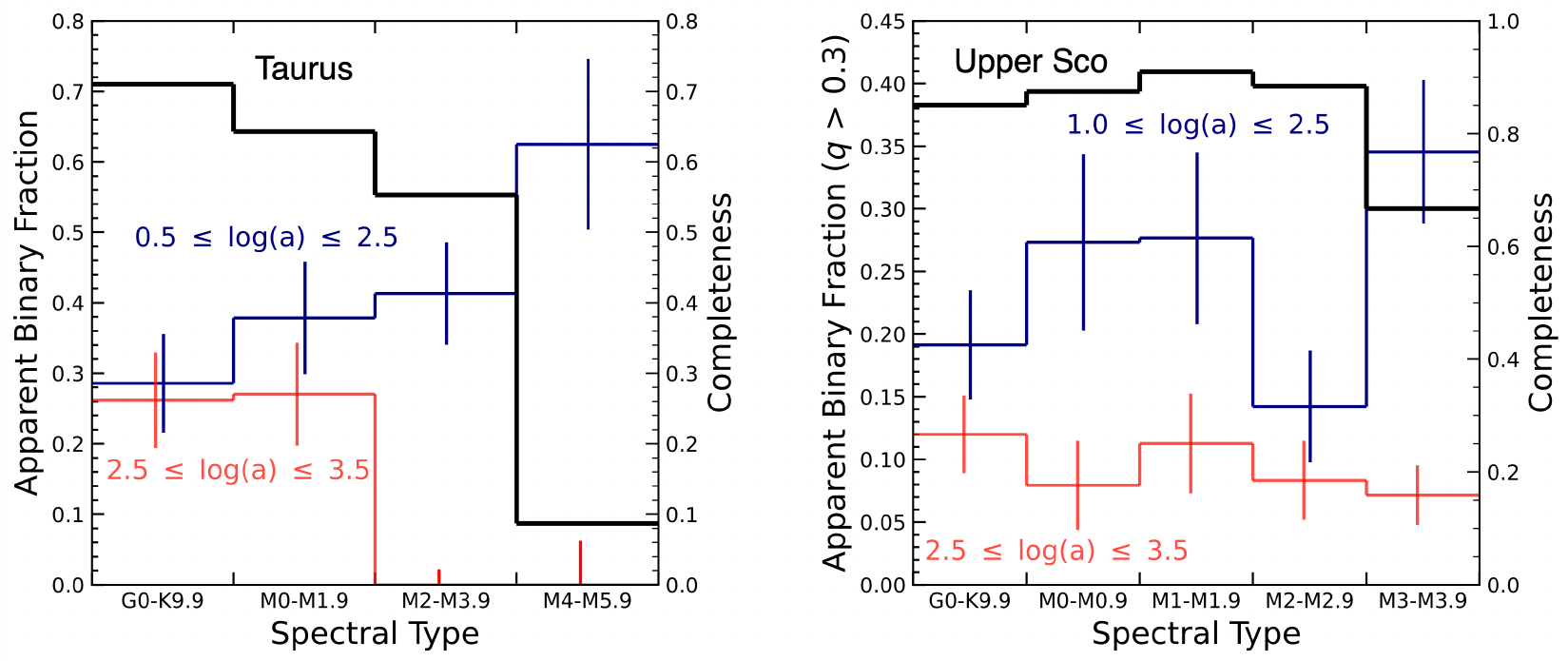}
    \caption{The apparent close binary fraction (blue) and apparent wide binary fraction (red) as a function of spectral type for Taurus (left panel) and Upper Sco (right panel). We also display the completeness (black), i.e., the fraction of Taurus and Upper Sco members targeted by the \citet{Kraus2011} and \citet{Tokovinin2020} surveys, respectively. The surveys are relatively complete and unbiased across earlier spectral types. Toward later spectral types, however, the completeness decreases and the magnitude-limited surveys become biased toward close binaries that have cleared out their dusty disks.}
    \label{fig:DiffSpec}
\end{figure*}

The distributions of G~magnitudes more fully illustrate the dichotomy. In the left panel of Fig.~\ref{fig:CDF}, we display the CDFs of the Taurus members with M2.5\,-\,M4.3 primaries separated into the same three categories as in Fig.~\ref{fig:MeanG}. Of the 13 close binaries, 12 are narrowly distributed across G = 12.3\,-\,14.3. The $\Delta$G~=~2.0~mag spread is consistent with the expected range of M2.5\,-\,M4.3 zero-age MS stars without any dust extinction \citep{Pecaut2013}. The single exception (MHO~2) is a moderately obscured G~=~17.0 close binary with a small mass ratio of $q$~=~0.34. Close extreme mass-ratio binaries may not clear out their disks as quickly as their twin binary counterparts. For example, compared to single-lined spectroscopic binaries (SB1s), \citet{Kounkel2019} showed that T~Tauri SB2s with $q$~$>$~0.6 exhibit a higher ratio of  Class III to Class II disks. It is thus not surprising that the close binary with $q$~=~0.34 in our subset still exhibits A$_{\rm G}$~=~3~mag of dust extinction. Meanwhile, both the single star / wide binary and unobserved subsets exhibit a much broader range of magnitudes, extending from G~=~12 to the {\it Gaia} detection limit of G~=~21. A Kolmogorov-Smirnov (KS) test reveals that the G~magnitude distributions of the close binary subset versus the single star / wide binary and unobserved subsets are discrepant at the $p_{\rm KS}$ = 0.05 (2.0$\sigma$) and $p_{\rm KS}$ = 0.003 (3.1$\sigma$) levels, respectively. A larger fraction of the single stars, wide binaries, and unobserved systems suffer from substantial dust extinction.

In the right panel of Fig.~\ref{fig:CDF}, we display the CDFs of G magnitudes for the 173 Upper Sco members with spectral types M3\,-\,3.9 separated into the same three categories. Both the 28 close binaries and 58 single stars / wide binaries span $\Delta$G = 2.0~mag, but are offset from each other, G = 12.8\,-\,14.8 and 13.1\,-\,15.1, respectively. The 87 unobserved members are substantially fainter, extending down to G = 18.4. Even after removing the most obscured member at G~=~18.4, the remaining $\Delta G$~=~4.4~mag spread across G = 12.8\,-\,17.2 signifies that many Upper Sco members near the \citet{Tokovinin2020} magnitude limit experience a broad range of dust extinction.

Considering that close binaries have preferentially cleared out their disks, we expect that the apparent close binary fraction will artificially increase toward later spectral types as we approach the magnitude limit of the imaging surveys. In the left panel of Fig.~\ref{fig:DiffSpec}, we display the apparent binary fractions across log\,$a$\,(au) = 0.5\,-\,2.5 and 2.5\,-\,3.5 as a function of primary spectral type for the Taurus members that \citet{Kraus2011} observed.  For each spectral type bin, we also display the fraction of Taurus members that \citet{Kraus2011} targeted with AO imaging in search of visual binaries. Across G0\,-\,M1.9, the close and wide binary fractions are relatively constant at 34\% and 26\%, respectively. The \citet{Kraus2011} survey is $\approx$\,70\% complete across this interval, and therefore the close and wide binary fractions are relatively unbiased. Toward later spectral types, however, the completeness drops, especially for the last bin where \citet{Kraus2011} targeted only 9\% of M4-M5.9 Taurus members. The wide binary fraction plummets to 0\% while the close binary fraction gradually increases, reaching 63\% for M4\,-\,M5.9 primaries. The M2-5.9 primaries are biased toward the bright, close binaries that have cleared out their disks and biased against the faint, highly obscured single stars and wide binaries. We therefore limit Taurus to G0\,-\,M1.9 members where the \citet{Kraus2011} survey is relatively complete and unbiased.

\begin{figure*}[t!]
    \centering
    \includegraphics[scale=0.65]{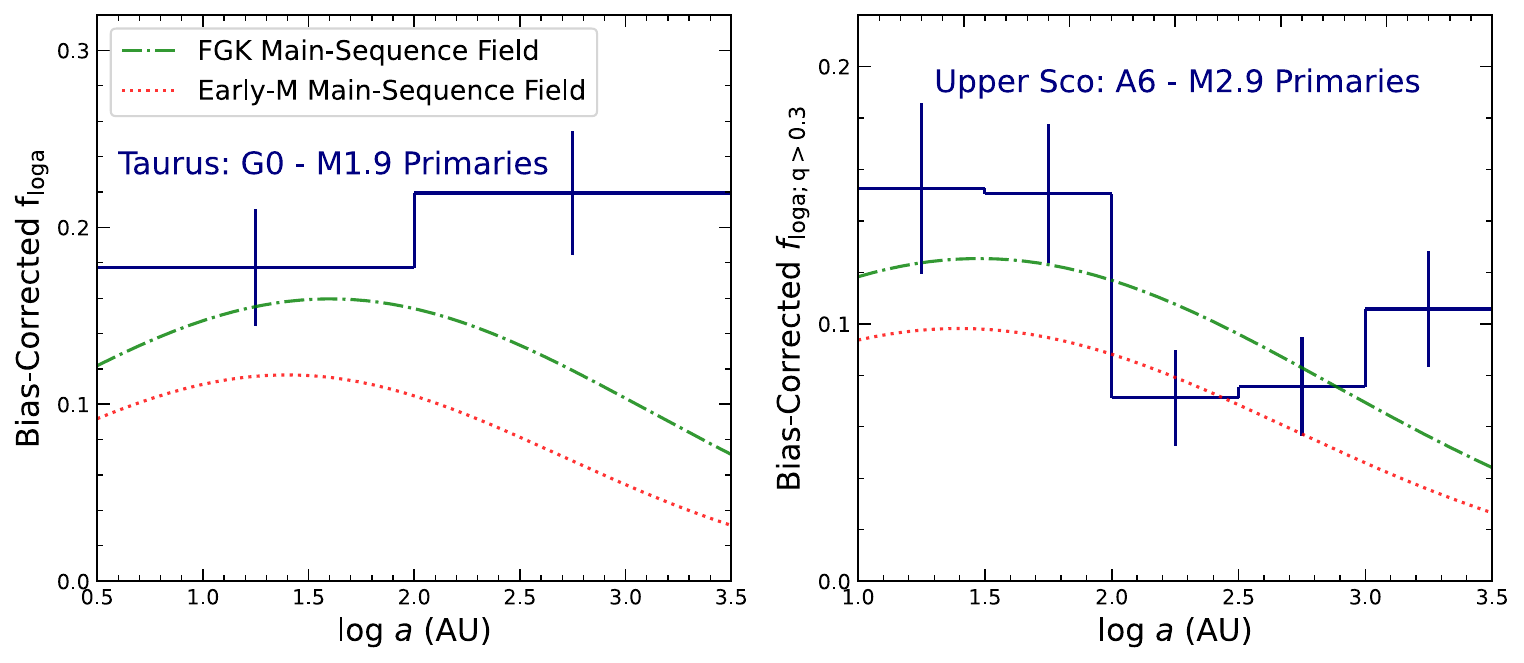}
    \caption{Similar to Fig.~\ref{fig:SepDist_NoCut} but now limited to G0\,-\,M1.9 Taurus members (left panel) and A6\,-\,M2.9 Upper Sco members (right panel) where the \citet{Kraus2011} and \citet{Tokovinin2020} surveys, respectively, are relatively complete and unbiased. The bias-corrected binary fractions below $a$~$<$~100~au in both Taurus and Upper Sco are now consistent with the field MS populations. Taurus instead exhibits a statistically significant excess beyond $a$~$>$~100 au.}
    \label{fig:SepDist_Cut}
\end{figure*}

In the right panel of Fig.~\ref{fig:DiffSpec}, we display the apparent close and wide binary fractions in Upper Sco as a function of spectral type. The binary fractions have been corrected for incompleteness down to $q$ = 0.3, but no corrections for dust extinction have been applied. Across spectral types G0-M2.9, the close and wide binary fractions are consistent with 20\% and 10\%, respectively. The \citet{Tokovinin2020} survey is highly complete at $\approx$\,90\% across this interval.  For M3-3.9 spectral types, however, the completeness drops to 66\%. In turn, the apparent close binary fraction increases substantially to 35\%\,$\pm$\,6\% and the apparent wide binary fraction decreases slightly to 7\%\,$\pm$\,2\%. This shift provides further evidence that the \citet{Tokovinin2020} sample is biased near the magnitude limit toward close binaries that have preferentially cleared out their dusty disks. We therefore limit the Upper Sco sample to A6\,-\,M2.9 spectral types where the \citet{Tokovinin2020} survey is relatively complete and unbiased.

\section{Bias-Corrected Results}
\label{sec:Results}

In the left panel of Fig~\ref{fig:SepDist_Cut}, we display the Taurus companion fraction $f_{\rm loga}$ per decade of orbital separation as in Fig.~\ref{fig:SepDist_NoCut}, but now limited to the 79 G0\,-\,M1.9 ($M_1$~=~0.51\,-\,2.5\,\Msun) primaries where the \citet{Kraus2011} survey is relatively complete and unbiased. We list the binary properties for this subset in Table~\ref{tab:Taurus}. The binary fraction of Taurus G0\,-\,M1.9 members across $a$~=~3\,-\,100~au is 27\%\,$\pm$\,5\%, which is slightly higher than but now fully consistent with the field MS value of 22\%. Meanwhile the Taurus G0\,-\,M1.9 binary fraction across $a$ = 100\,-\,3,000~au is 33\%\,$\pm$\,5\%, corresponding to a factor of 2.1 excess that is statistically significant at the 3.3$\sigma$ level. By accounting for the dust-extinction selection bias and limiting our sample to spectral type instead of magnitude, the apparent binary excess below $a$~$<$~100~au disappears while the binary excess beyond $a$~$>$~100~au becomes statistically significant.

In the right panel of Fig.~\ref{fig:SepDist_Cut}, we again display the completeness-corrected separation distribution $f_{\mathrm{loga;q>0.3}}$ of binaries with $q$~$>$~0.3 in Upper Sco, but this time limited to the 397 members with primary spectral types A6\,-\,M2.9 for which the \citet{Tokovinin2020} sample is relatively complete and unbiased. We report the bias-corrected binary properties for this subset in Table~\ref{tab:UpperSco}.  The completeness-corrected binary fraction for Upper Sco A6\,-\,M2.9 members across $a$~=~10\,-\,100~au and above $q$ $>$ 0.3 is 15\%\,$\pm$\,2\%, fully consistent with the solar-type MS field value of 13\%. As with Taurus, the apparent excess of binaries across $a$~=~10\,-\,100\,au in Upper Sco disappears when we limit the \citet{Tokovinin2020} sample to spectral type instead of magnitude. 

 Across log\,$a$\,(au) = 2.0\,-\,3.5, the bias-corrected Upper Sco measurements are consistent with the field population. There appears to be a slight excess of binaries across log\,$a$\,(au) = 3.0\,-\,3.5 compared to the log-normal separation distribution, but the excess is only discrepant at the 1.9$\sigma$ level. Moreover, the observed field population of FGK-dwarf binaries exhibits a similar excess near log\,$a$\,(au) = 3.2 compared to the simplistic log-normal separation distribution (see Fig.~\ref{fig:All_Before}). The overall bias-corrected binary fraction in Upper Sco across log\,$a$ = 2.0\,-\,3.5~au and above $q$~$>$~0.3 is 13\%\,$\pm$\,2\%, which precisely matches the field value of 13\%. The wide binary fraction in Upper Sco matches the field value.

Finally, we account for dust extinction in the \citet{Duchene2018} sample of ONC binaries. Due to the extremely young age of the ONC and the stochastic nature of accretion, spectral type does not directly correspond to mass. We therefore adopt the stellar masses presented in Table~1 of \citet{Duchene2018}, which derived mostly from SED fitting \citep{Hillenbrand1997,DaRio2010}. To account for the dust-extinction bias near the magnitude-limit, we remove the 16 low-mass systems with $M_1$~=~0.37\,-\,0.7\,\Msun, including three tight binaries. We also remove the six systems above $M_1$~$>$~1.6\,\Msun\ that will evolve into A-dwarfs, including two tight binaries. We list the binary properties of the ONC members with $M_1$ = 0.7\,-\,1.6\,\Msun\ in Table~\ref{tab:Orion}. The binary fraction across $a$~=~10\,-\,60\,au of ONC stars with $M_1$ = 0.7\,-\,1.6\,\Msun\ is 19$_{-7}^{+12}$\%, smaller than before and now fully consistent with the FGK MS field value of 13\%. By narrowing the \citet{Duchene2018} sample to include only primaries with $M_1$ = 0.7\,-\,1.6\,\Msun\ that will evolve into FGK dwarfs, the ONC binary fraction across $a$~=~10\,-\,60~au is no longer discrepant with the corresponding FGK MS field population.

\section{Comparison of Different SFRs}
\label{sec:Comparison}
\begin{figure}[t!]
    \centering
    \includegraphics[trim = 3mm 5mm 3mm 5mm, width=3.4in]{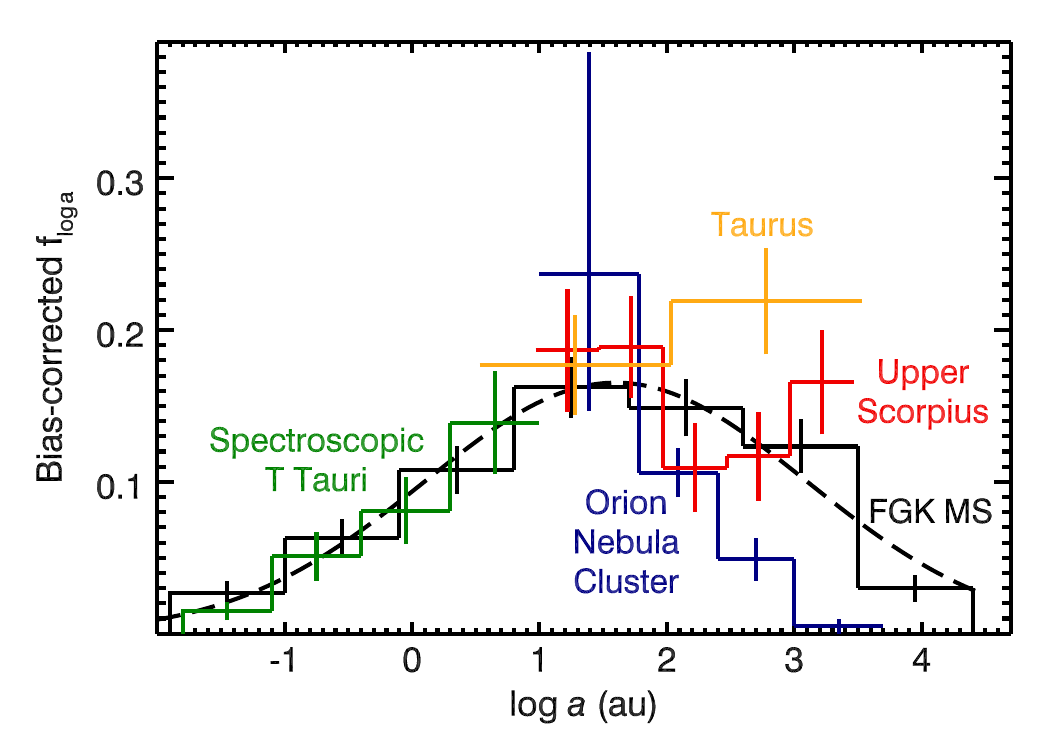}
    \caption{Similar to Fig.~\ref{fig:All_Before}, but now limiting the surveys for visual binaries according to primary spectral type / mass, which mitigates the dust-extinction selection bias. Across $a$ = 10\,-\,100 au, all three SFRs are now consistent with the field MS binary population, and the separation distribution between the spectroscopic and imaging surveys are now continuous. Beyond $a$~$>$~100~au, low-density Taurus shows a real, statistically significant excess (mostly tertiaries) while the dense ONC exhibits a deficit due to dynamical disruptions.}
    \label{fig:All_After}
\end{figure}

We have re-examined the properties and occurrence rates of visual binaries in Taurus, Upper Sco, and the ONC. All three environments exhibit a broad range of dust extinction from circumstellar disks that affects single stars and wide binaries more severely than close binaries that have preferentially truncated their disks. In Fig.~\ref{fig:All_After}, we display our bias-corrected binary separation distributions for all three environments. For Taurus, we show the same two data points from the left panel of Fig.~\ref{fig:SepDist_Cut}. For the ONC, we display our bias-corrected measurement across $a$ = 10\,-\,60 au from Table~\ref{tab:Orion} and the relatively unbiased measurements beyond $a$~$>$~60~au from \citet{Scally1999} and \citet{Reipurth2007}.

For Upper Sco, we must first convert $f_{\rm loga;q>0.3}$ into $f_{\rm loga}$ by correcting for incompleteness down to $M_2$ = 0.08\,\Msun. We adopt the separation-dependent mass-ratio distributions for solar-type binaries from \citet{Moe2017}. Across $a$~=~10\,-\,100~au, solar-type binaries follow a uniform mass-ratio distribution with a 10\% excess of twins above $q$~$>$~0.95. This distribution results in a completeness ratio of $f_{\rm loga;q>0.3}$/$f_{\rm loga}$ = 0.81. Meanwhile, near $a$~=~1,000~au, solar-type binaries are skewed toward smaller mass ratios: $f_q$~$\propto$~$q^{-1}$ across $q$ = 0.3\,-\,1.0, a uniform distribution below $q$ $<$ 0.3, and a negligible excess twin fraction. This mass-ratio distribution yields a ratio of $f_{\rm loga;q>0.3}$/$f_{\rm loga}$ = 0.64. We apply the separation-dependent completeness corrections and display in Fig.~\ref{fig:All_After} the results for $f_{\rm loga}$ in Upper Sco. 

By limiting the previously observed samples to primary spectral type / mass, the apparent excess across $a$ = 10\,-\,100~au disappears in all three environments. The occurrence rate of T Tauri binaries across intermediate separations is now fully consistent with the field MS population. The T~Tauri binary separation distribution is now also continuous between spectroscopic binaries ($a$~$<$~10~au) and visual binaries ($a$ $>$ 10~au). 

Although the binary properties below $a$~$<$~100~au in the three SFRs are now consistent with the field MS population, the companion fraction beyond $a$~$>$~100~au depends on environment. At these wide separations, Taurus exhibits a statistically significant excess of companions, the ONC shows a deficit, and Upper Sco roughly matches the solar-type field MS population. We conclude that field MS stars derived from an amalgam of various SFRs similar to Taurus, Upper Sco, and the ONC. However, the representative cluster is no longer the ONC as previously argued, but instead Upper~Sco corresponds to the average birth environment of solar-type stars. For every solar-type star born in a dense environment like the ONC that exhibits a deficit of companions beyond $a$~$>$~100~au due to dynamical disruptions, there must be a corresponding solar-type star born in a sparse association like Taurus that displays an excess of wide companions. 

\section{Implications for N-body Interactions}
\label{sec:Implications}

Close binaries within $a$~$<$~5~au cannot form in~situ \citep{Boss1986, Bate1998}. Instead, companions initially fragment on larger scales and migrate inward  via interactions with the surrounding gas \citep{Bate1995,Bate1997,Bate2002,Moe2018,TokovininMoe2020}. Even after the gas dissipates, N-body interactions with other stars in their birth cluster continue to alter the binary separation distribution. Hard (close) binaries tend to harden while soft (wide) binaries tend to soften, possibly becoming dynamically disrupted \citep{Heggie1975}. Numerical simulations of young binaries in clusters demonstrate that the wide binary fraction was initially higher before a subset was dynamically disrupted via N-body interactions
\citep{Kroupa1995,Fregeau2004,CournoyerCloutier2024}. The gas remaining during the Class II/III T~Tauri phase is insufficient to alter binary orbits via dynamical friction. Differences between the separation distributions of Class~II/III T~Tauri binaries versus the field population must therefore be a result of N-body interactions. 

\begin{figure}[t!]
    \centering
    \includegraphics[trim = 3mm 5mm 3mm 5mm, width=3.4in]{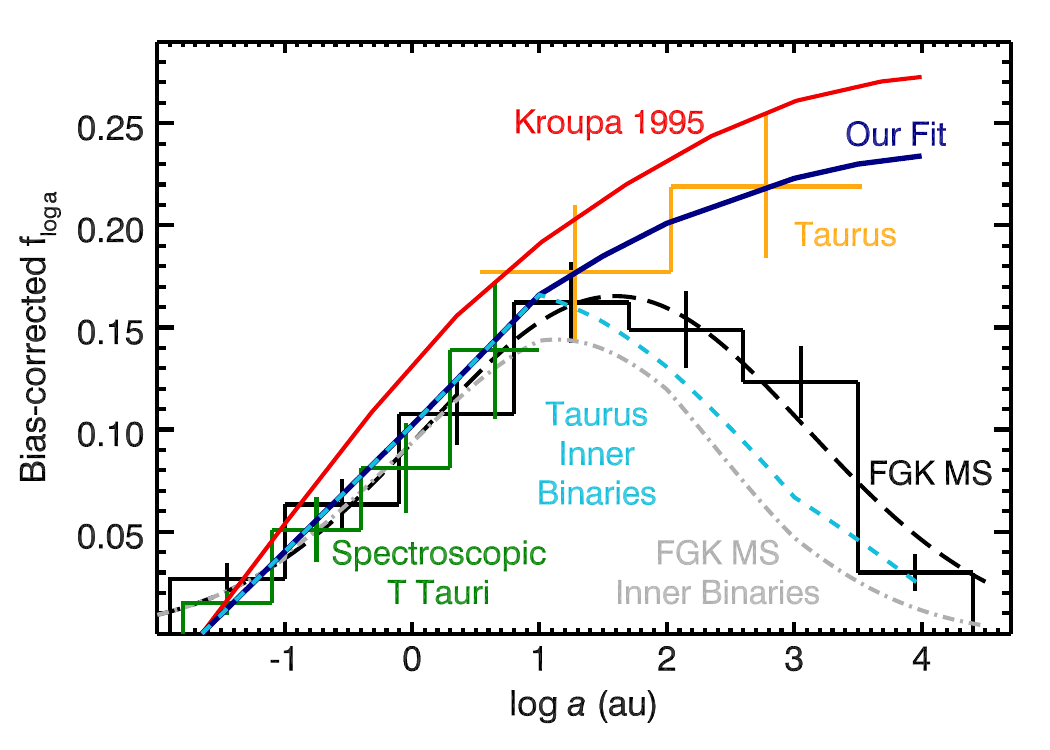}
    \caption{Similar to Fig.~\ref{fig:All_After}, but emphasizing the differences between our bias-corrected Taurus measurements (orange) and the field solar-type MS population (black). The previous fit to T~Tauri binaries in low-density SFRs \citep[][red]{Kroupa1995} was overestimated due to the dust-extinction selection bias of previous imaging surveys. Our fit (dark blue) is anchored to the bias-corrected Taurus measurements. The inner binary separation distributions of field MS stars (dot-dashed grey) and Taurus members (dashed light blue) are lower than their overall companion distributions. Surprisingly, the total binary fraction of 52\%\,$\pm$\,7\% within $a$~$<$~10,000~au in Taurus is quite similar to the field value of 45\%. The excess of wide companions in Taurus are mostly outer tertiaries in hierarchical triples.}
    \label{fig:floga_Taurus}
\end{figure}

The sparse Taurus association provides an ideal laboratory for measuring the primordial properties of multiple stars with minimal N-body interactions. In Fig.~\ref{fig:floga_Taurus}, we compare our bias-corrected Taurus measurements alongside the field FGK MS separation distribution. \citet{Kroupa1995} previously fit a T~Tauri binary period distribution based on the observations to low-density SFRs available at that time (see their Section~5 and bottom panel of their Fig.~8). We display their T~Tauri separation distribution as the red line in our Fig.~\ref{fig:floga_Taurus}. \citet{Kroupa1995} suggested that all stars are born in binaries, i.e., an initial binary fraction of 100\%, and that dynamical interactions in dense clusters subsequently disrupt the widest systems. By construction, integrating the red curve in our Fig.~\ref{fig:floga_Taurus} yields a 100\% binary fraction below $a$~$<$~10,000~au. We scale the \citet{Kroupa1995} model to our bias-corrected measurements for Taurus, resulting in the dark blue curve in Fig.~\ref{fig:floga_Taurus}. Integrating our fit yields a total companion frequency of CF = 0.84\,$\pm$\,0.07, which is smaller than unity but still larger than the field FGK MS value of CF = 0.60.  

Our fit across $a$~=~0.1\,-\,10~au is only 9\% higher than the field MS value (percentage difference between dark blue and dashed black curves in Fig.~\ref{fig:floga_Taurus}). As discussed in \citet{Kounkel2019}, the bias-corrected spectroscopic binary fraction and separation distribution within $a$~$<$~10~au match the field MS population with no statistically significant trend with respect to SFR density or age. A 9\% difference between low-density SFRs and the field are within the uncertainties. The measured offset suggests that only 9\% of pre-MS systems averaged across all environments lose a wide solar-type companion due to dynamical disruptions. The disruption of solar-type companions increases the total number of solar-type single stars in a population, which effectively reduces the close binary fraction by a factor of 9\%. This 9\% reduction is less significant than the 35\% effect previously estimated, i.e., the percentage difference between the red and dashed black curves across $a$ = 0.1\,-\,10~au in Fig.~\ref{fig:floga_Taurus}.

Toward wider separations, we must distinguish inner binaries from outer tertiaries in hierarchical triples. For solar-type MS field stars, we adopt the inner binary separation distribution from \citet[][dash-dot grey line in our Fig.~\ref{fig:floga_Taurus}]{Moe2021}. In the field, most companions within $a$ $<$ 10 au are inner binaries while three quarters of companions near $a$~=~10,000~au are outer tertiaries. Integrating the dash-dot grey line in our Fig.~\ref{fig:floga_Taurus} recovers the solar-type MS field binary fraction of BF = 46\% \citep{Raghavan2010,Tokovinin2014}. Limiting the integral to $a$~$<$~10,000~au yields a marginally smaller binary fraction of BF = 45\%. 

The triple star fraction in Taurus is substantially higher than in the field. For the unbiased subset of 79 G0-M1.9 Taurus members, \citet{Kraus2011} resolved 22 companions across log\,$a$\,(au) = 2.5\,-\,3.5, of which 13 (59\%\,$\pm$\,10\%) are outer tertiaries where the inner binaries were also spatially resolved beyond $a_{\rm in}$~$>$~3~au. This tertiary fraction is a lower limit considering some additional companions across log\,$a$\,(au) = 2.5\,-\,3.5 may be outer tertiaries where the inner binaries remain unresolved below $a_{\rm in}$~$<$~3~au. The triple star fraction in Taurus increases substantially for companions beyond $a$ $>$ 1,000 au. In our G0-M1.9 subset, \citet{Kraus2011} identified 11 companions across log\,$a$\,(au) = 3.0\,-\,3.6, of which 10 (91\%\,$\pm$\,9\%) are tertiaries. We adopt a simple piecewise model for the Taurus inner binary probability function $p_{\rm in}(a)$ that matches these observations. We assume that $p_{\rm in}$ = 100\% of companions below $a$~$<$~10~au are inner binaries, the inner binary probability decreases to $p_{\rm in}$ = 30\%\,$\pm$\,10\% at $a$ = 1,000 au, and that only $p_{\rm in}$ = 10\%\,$\pm$\,10\% of companions at $a$~=~10,000~au are inner binaries. We linearly interpolate these fractions with respect to log\,$a$.

Multiplying the overall Taurus companion separation distribution $f_{\rm loga}$($a$) (our dark blue fit in Fig.~\ref{fig:floga_Taurus}) by our piecewise model for the inner binary probability $p_{\rm in}$($a$) yields the inner binary separation distribution in Taurus:

\begin{equation}
 f_{\rm loga;in} = f_{\rm loga}\,p_{\rm in} .
\end{equation}

\noindent We display $f_{\rm loga;in}$($a$) in Taurus as the light blue dashed curve in Fig.~\ref{fig:floga_Taurus}. Remarkably, the inner binary separation distribution in Taurus nearly matches its field MS counterpart. Integrating the dashed light blue curve results in a total binary fraction of BF = 52\%\,$\pm$\,7\% in Taurus, where the uncertainty derives from propagating the errors in both our fit to the Taurus separation distribution and our model for the inner binary probabilities. The binary fraction in Taurus is only marginally higher than the field MS value of 45\% within the same separation range and consistent with each other at the 1.0$\sigma$ level. The difference between the overall companion distribution in Taurus (dark blue fit in Fig.~\ref{fig:floga_Taurus}) and the field (dashed black) can be explained almost exclusively due to the enhanced triple star fraction in Taurus. Ignoring the negligible contribution from quadruples, the triple fraction in Taurus is TF = CF\,$-$\,BF = 0.84\,$-$\,0.52 = 32\%\,$\pm$\,5\%, which is more than double the field MS value of TF = 0.60\,-\,0.46 = 14\%. 

We consider a hypothetical scenario where Taurus members are placed in a denser environment like Upper~Sco where N-body interactions and disruptions could potentially yield the field MS population. A Taurus solar-type star would need to lose 0.1\,-\,0.2 companions per primary on average to match the field companion frequency, but such disrupted companions must be predominantly tertiaries. About 90\% of the companions beyond $a$~$>$~1,000~au in Taurus are outer tertiaries, and so it is not surprising that dynamical processing in denser clusters would preferentially disrupt the tertiaries. Moreover, most wide companions are low-mass stars that will evolve into M-dwarfs. In our unbiased subset of 79 G0-M1.9 primaries in Taurus, \citet{Kraus2011} resolved 25 companions across log\,$a$ (au) = 2.5\,-\,3.6, of which only 11 are solar-type stars with $M_{\rm comp}$ = 0.7\,-\,1.2\,\Msun. The remaining 14 companions have $M_{\rm comp}$ $<$ 0.7\,\Msun. If all 11 wide solar-type companions were disrupted, then there would be 11 more solar-type single stars and the close binary fraction would be reduced by a factor of 11/79 = 14\%. In reality, some wide solar-type companions will remain gravitationally bound, as demonstrated by the existence of wide solar-type companions in the field. For consistency sake, we assume that 7 of the 11 wide solar-type companions, all tertiaries, would become dynamically disrupted, which would reduce the close binary fraction by the factor of 7/79 = 9\% we estimated above. The total binary fraction would reduce to BF = 52\%/1.09 = 48\%. N-body interactions would therefore need to disrupt only two inner binaries to further reduce the binary fraction by 2/79 = 3\% to the field value of 45\%. 

Compared to the field MS population, we conclude that Taurus exhibits a significantly enhanced triple star fraction and only a slightly enhanced binary fraction. If Taurus members were embedded in a cluster with average density, N-body interactions would predominantly disrupt the outer tertiary companions, including some solar-type companions. Inner binaries at systematically shorter separations are less prone to disruption, and N-body interactions with inner binaries below $a$~$<$~100~au would be negligible. 

\section{Summary}
\label{sec:Summary}

We summarize our main results as follows:

\begin{itemize}
\item For class II/III T~Tauri stars, circumstellar disks cause A$_{\rm G}$~=~2\,-\,7~mag of dust extinction. Close binaries within $a$~$<$~100~au clear out and truncate their disks on faster timescales compared to single stars or wide binaries. For a narrow range of spectral types, we showed that close binaries are systematically brighter than single stars and wide binaries due to less dust extinction. The magnitude-limited samples of previous AO/speckle imaging surveys are therefore biased toward close binaries within $a$~$<$~100~au that have cleared out their dusty disks. 

\item By limiting the previously observed samples of T~Tauri stars in Taurus, Upper Sco, and the ONC according to primary spectral type / mass instead of magnitude, the apparent excess of binaries across $a$ = 10\,-\,100~au disappears in all three environments. The occurrence rate of T~Tauri binaries across intermediate separations is now fully consistent with the field MS population. The T~Tauri binary separation distribution is now continuous between spectroscopic binaries ($a$~$<$~10~au) and visual binaries ($a$ $>$ 10~au). 

\item Beyond $a$~$>$~100~au, sparse Taurus exhibits an excess of companions (mostly outer tertiaries), the dense ONC displays a deficit due to dynamical disruptions, and Upper Sco roughly matches the field MS population. Upper Sco therefore corresponds to the average birth environment of solar-type stars. 

\item The triple star fraction in Taurus is TF = 32\%\,$\pm$\,5\%, which is more than double the field MS value of TF = 14\%. Within $a$~$<$~10,000~au, the binary star fraction in Taurus is BF = 52\%\,$\pm$\,7\%, which is only slightly larger than and consistent with the field MS value of BF = 45\%. 

\item If Taurus members were embedded in a birth environment of average density, then predominantly outer tertiaries beyond $a$~$>$~1,000~au would become dynamically disrupted. Inner binaries at systematically shorter separations are less prone to disruption, and N-body interactions with inner binaries below $a$~$<$~100~au would be negligible. 

\end{itemize}

This work was funded by Wyoming NASA Space Grant Consortium, NASA Grant \#80NSSC20M0113.

\bibliographystyle{aasjournalv7}                       
\bibliography{moe_biblio}

\end{document}